 \documentclass[prb,twocolumn,showpacs,amsmath,amssymb,
                letterpaper,citeautoscript]{revtex4}

\usepackage[dvips]{graphicx}% Include figure files
\usepackage{dcolumn}% Align table columns on decimal point
\usepackage{bm}% bold math
                                                                                                              
\begin{document}
 
\title{
Molecular Conductance:
Chemical Trends of Anchoring Groups
}
 
\author{
San-Huang Ke,$^{\dag,\ddag}$ Harold U. Baranger,$^{\ddag}$ and Weitao Yang$^{\dag}$
}
 
\affiliation{
$^{\rm \dag}$Department of Chemistry, Duke University, Durham, NC 27708-0354 \\
$^{\rm \ddag}$Department of Physics, Duke University, Durham, NC 27708-0305
}
 
\date{\today}
 
\begin{abstract}
Combining density functional theory calculations for molecular electronic structure with a Green function method for electron transport, we calculate from first principles the molecular conductance of benzene connected to two Au leads through different anchoring atoms -- S, Se, and Te. The relaxed atomic structure of the contact, different lead orientations, and different adsorption sites are fully considered.  We find that the molecule-lead coupling, electron transfer, and conductance all depend strongly on the adsorption site, lead orientation, and local contact atomic configuration.  For flat contacts the conductance decreases as the atomic number of the anchoring atom increases, regardless of the adsorption site, lead orientation, or bias.  For small bias this chemical trend is, however, dependent on the contact atomic configuration: an additional Au atom at the contact with the (111) lead changes the best anchoring atom from S to Se although for large bias the original chemical trend is recovered.
\end{abstract}

\maketitle

\section{Introduction}

A critical issue in molecular electronics \cite{mol2,mol3,mol4,mol5} is to find anchoring groups and construct contact structures which provide both stability and high contact transparency.  In many recent experiments \cite{bj1,bj2,bj3,xiao,khondaker}, Au electrodes were used as leads for electronic current because of their high conductivity, stability, and well-defined fabrication technique.  A common way to construct a lead-molecule-lead (LML) system is by using a break junction, formed either mechanically \cite{bj1,bj2,bj3,xiao,h2} or electrically \cite{khondaker,park,morpurgo,fuhrer}. In these break-junction experiments, the atomic structure of the molecule-lead contact is unknown.  Therefore, neither the influence of detailed atomic structure on transport through the molecules nor a path to improved performance is clear.  With regard to anchoring groups, a S atom is commonly used \cite{mol3,bj1,bj2,bj3,xiao,khondaker} to connect various benzene-like organic molecules to Au leads because of its good binding with Au and good stability. However, a recent experiment for $\alpha$,$\omega$-bisacetylthio-terthiophene and $\alpha$,$\omega$-bisacetylseleno-terthiophene \cite{patrone} showed that a Se anchoring atom is better than S for zero bias conductance. Currently, it is an open problem whether there are other better choices and what the system dependence is.

In terms of theoretical studies, there are mainly two {\it ab initio} approaches for electron transport through molecules.  One was developed by Lang, {\it et al.} \cite{jellium}: the Kohn-Sham equation of the system is mapped into a Lippmann-Schwinger scattering equation which is solved for the scattering states self-consistently.  In the implementation \cite{jellium,jellium2}, the jellium model was adopted for the two metallic electrodes of an LML system.  The other approach \cite{datta1,datta2,pal1,mcdcal,transiesta,trank1}, is based on a density functional theory (DFT) calculation for the molecular electronic structure combined with a non-equilibrium Green function (NEGF) method for electron transport. Very close to the latter approach there is also an approach based on a self-consistent tight-binding method combined with the NEGF method \cite{mingo1,mingo2}.

Within the DFT+NEGF method itself, there are generally two approaches describing the boundary between the leads and the molecule of a LML system. In one category, a \textit{cluster} geometry is adopted, either for all the subsystems of a LML system or for only the extended molecule while the leads are handled by an extended system method (for example, Refs.  \cite{datta1,datta2,pal1,xue1,xue2,non-sc}).  
%An important advantage in using clusters is that it is then convenient to employ well-established quantum chemistry codes to do the electronic structure calculations.  In order to fully accommodate strong molecule-lead coupling, however, it is necessary to include large parts of the leads in the extended molecule. This introduces artificial surface effects; in order to eliminate them a very large system is needed. As this is difficult in practice, often only several lead atoms (six is a typical number) are included with the molecule (for example, Refs. \cite{xue1,xue2,non-sc}).  In this case, significant artificial surface effects are inevitable, the contact atomic relaxation cannot be included, and so accurate molecule-lead coupling is not available. In the case that the extended lead is treated with a tight-binding description, artificial scattering at the interface between the tight-binding part of the lead and the DFT part of the lead will, of course, occur although it may not be significant if many lead atoms are included in the extended molecule.
The other type of DFT+NEGF approach \cite{mcdcal,transiesta,trank1} uses \textit{periodic boundary conditions} (PBC) with large parts of the leads included in the extended molecule, so that the interaction between the molecule and its images will be screened off by the metallic lead in between. In this case all the subsystems can be treated on the same footing. 
%In this case all the potential problems mentioned above will be absent, the whole LML system becomes nearly perfect in geometry, and all the subsystems are treated on exactly the same footing.

Concerning the important issue of the best anchoring group, surprisingly few studies have been done.  First, there is a theoretical calculation \cite{jellium,jellium2} for the benzene molecule using jellium leads which demonstrated that the best anchoring atom for contact transparency is not an S atom but rather a Te atom -- it increases the equilibrium conductance by about 25 times. A potential problem with this calculation, however, is the neglect of the detailed electronic structure of Au as well as the contact atomic structure. Although in the DFT+NEGF method both molecule-lead interaction and contact structure can be taken into account in principle, in practice this has not been done: contact atomic structures were predetermined in all previous calculations.
%(for instance, Refs. \cite{datta1,datta2,pal1,mcdcal,transiesta,xue1,xue2,non-sc}). 
Thus, to the best of our knowledge, the effect of relaxing atomic positions in the contacts has not been investigated theoretically.  Very recently, the effect of three different molecular end groups has been evaluated with DFT+NEGF \cite{xue3}, though not the ones considered here nor with contact structure relaxation.

In this paper we investigate systematically the effects of different realistic contact atomic structures and re-examine the important issue of the best anchoring group.  We use our previously developed self-consistent implementation of the DFT+NEGF method \cite{trank1}.  In our method, as in the other two implementations \cite{mcdcal, transiesta}, PBC are adopted and large parts of the two metallic leads are included in the device region (about 40 Au atoms per lead) so that the molecule-lead interactions (including electron transfer and atomic relaxation) are fully included, and the electronic structure of the molecule and leads are treated on exactly the same footing.
%without introducing any artificial surface effects or scattering.

We calculate the molecular conductance of benzene connected to two Au leads with real atomic structure and finite width in order to simulate physical leads in break-junction experiments.  The connection is made through three different anchoring atoms: S, Se, and Te.  We consider fully the atomic relaxation of these LML systems, and consider different lead orientations and different adsorption sites of the anchoring groups.  The fully optimized atomic structures and high level basis set for all the atomic species make our calculations well converged.
%and more realistic. 
It is found that the molecule-lead coupling, the electron transfer, and therefore the conductance depend strongly on the adsorption site, lead orientation, and local contact atomic configuration.  It turns out that for ideally flat contacts the equilibrium conductance decreases with increasing atomic number of the anchoring group, regardless of the contact atomic structure and bias.  However, for contacts with a fluctuation of one Au atom, Se becomes slightly better than S for small bias, although for large bias the chemical trend goes back and S becomes better. This shows the critical role of the real contact atomic structure in electron transport through molecules.

\section{Computational Methods}

%=================================== Fig. 1 ========================================
\begin{figure}[htb]
\label{fig1}
\includegraphics[angle=  0,width= 6.8cm]{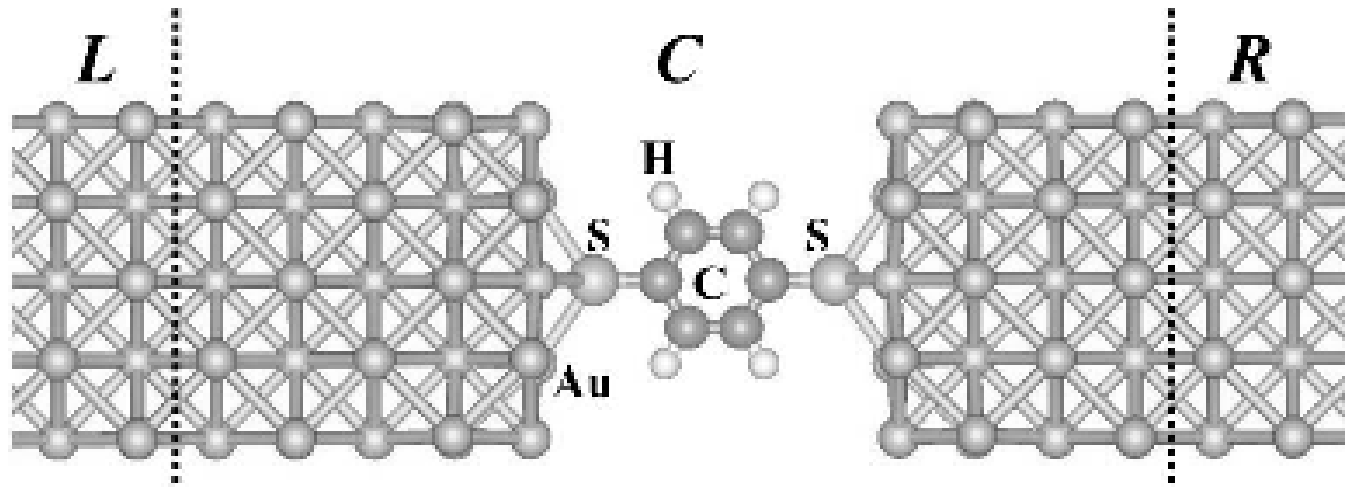} {\bf (a)}
\includegraphics[angle=  0,width= 7.2cm]{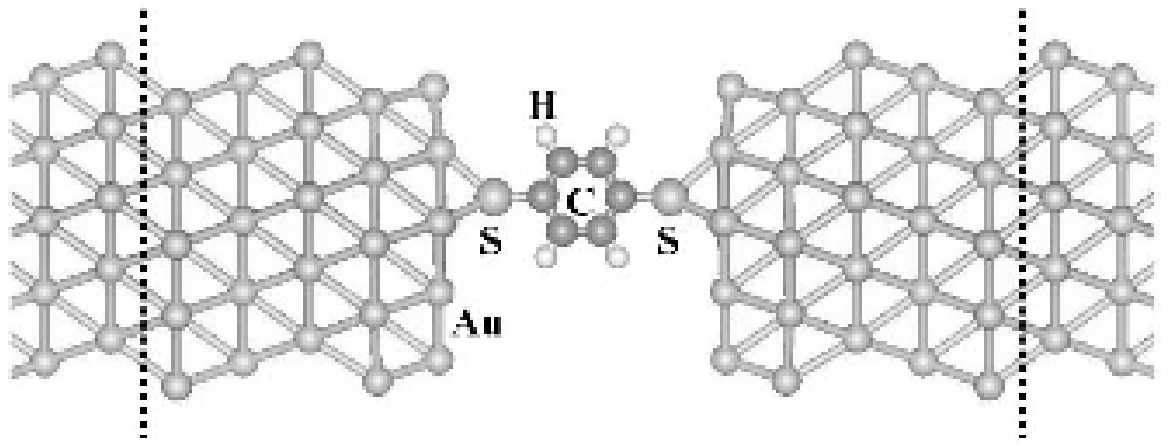} {\bf (b)} 
\includegraphics[angle=  0,width= 7.2cm]{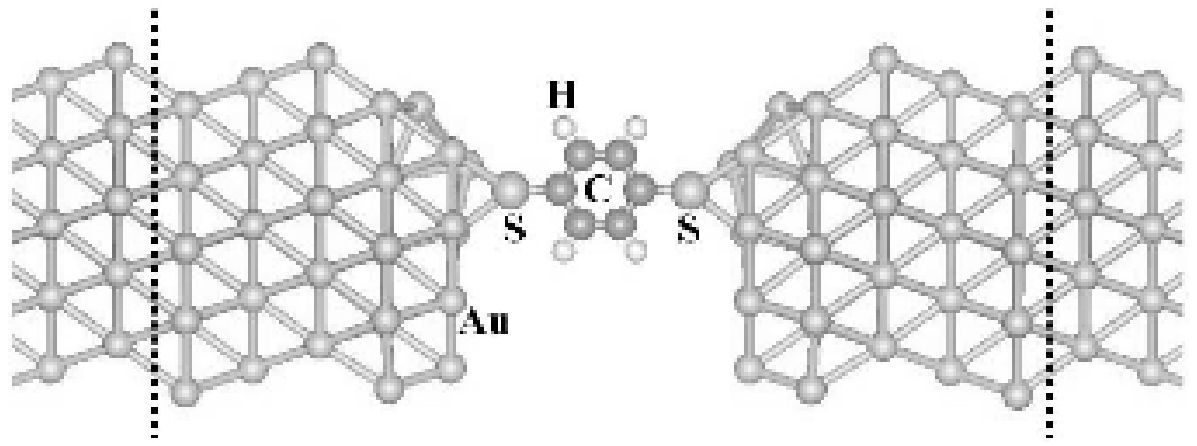} {\bf (c)}
\includegraphics[angle=  0,width= 7.2cm]{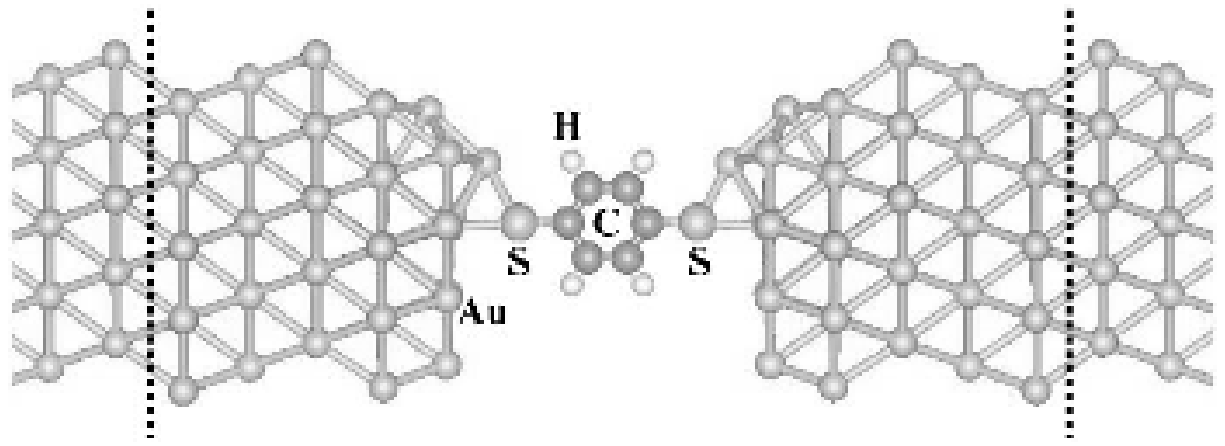} {\bf (d)}
\includegraphics[angle=  0,width= 1.7cm]{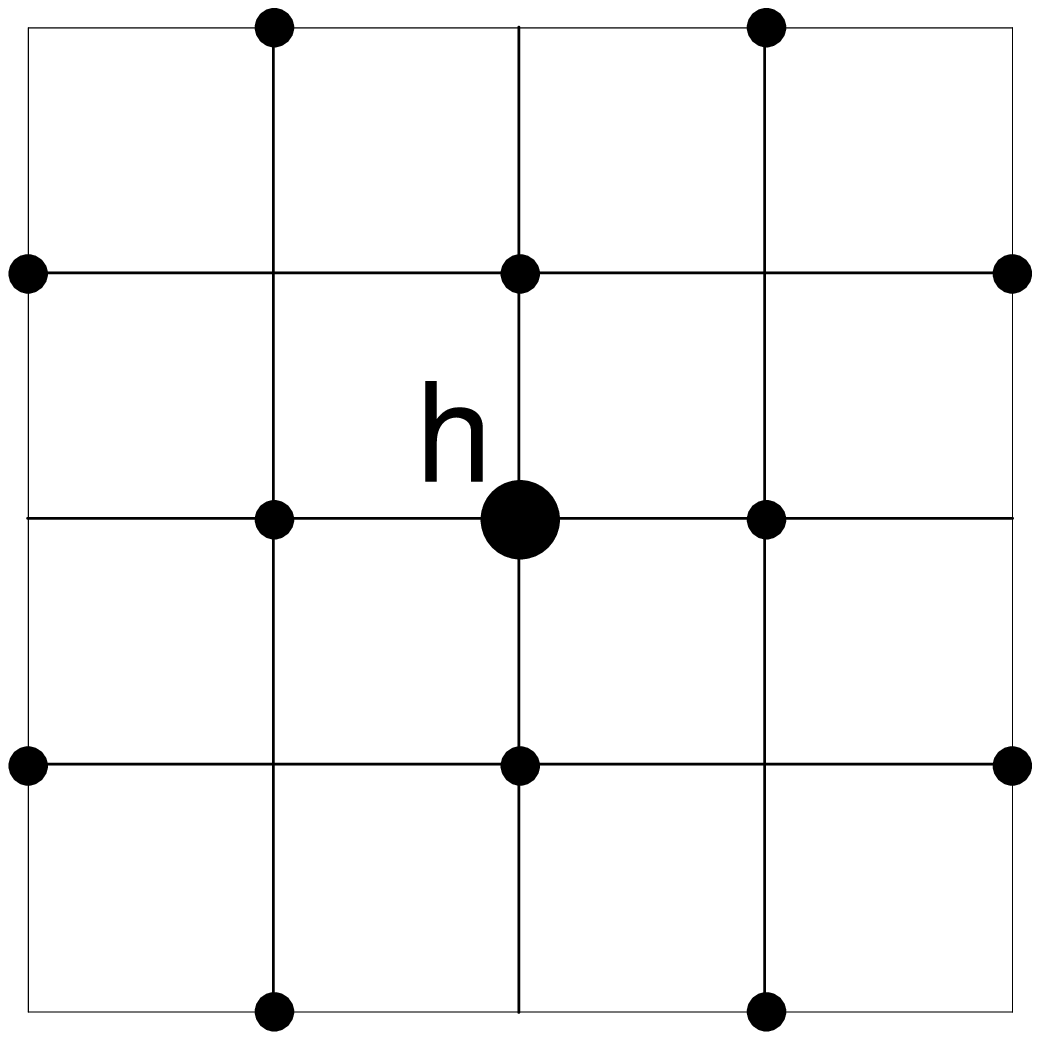} {\bf (e)}
\includegraphics[angle=  0,width= 1.5cm]{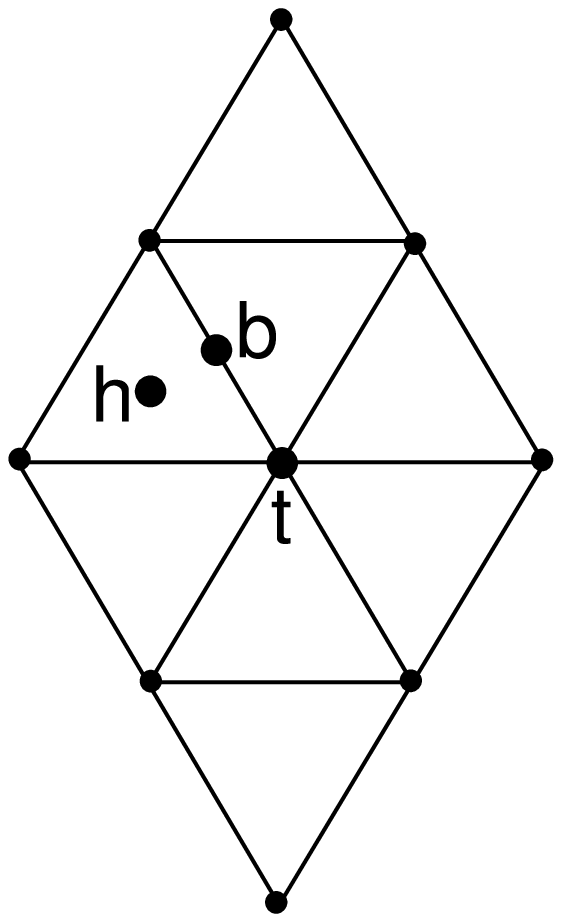} {\bf (f)}
\label{fig_str}
\caption{Optimized atomic structures for systems with two S anchoring atoms.  Optimized structures with two Se or Te anchoring atoms are similar and therefore not shown.  (a)~The Au leads are in the (001) direction, and the anchoring atom is located at the hollow site of the Au(001) surface [as shown in (e)]. (b)-(d) The Au leads are in the (111) direction, and the anchoring atom is located at the hollow, bridge, and top sites, respectively [as shown in (f)].  The dashed line denotes the interface between the device region ($C$) and the left or right lead ($L$ or $R$).  
} 
\end{figure}

%============================== Table 1 ========================
\begin{table*}[tbh]
\label{table1}
\caption{Binding energies between the lead and the
molecule (in eV), in terms of the total energies ($E_{tot}$'s)
of the whole and the subunits of a system.
For a relaxed or unrelaxed system it is determined by
$E_{tot}$(lead+molecule) -
$E_{tot}$(lead) - $E_{tot}$(molecule).
}
\begin{ruledtabular}
\begin{tabular}{cccccccc}
lead & adsorp. & \multicolumn{2}{c}{S-anchored} & \multicolumn{2}{c}{Se-anchored}
                  & \multicolumn{2}{c}{Te-anchored} \\
orient. & site & relaxed & unrelaxed & relaxed & unrelaxed & relaxed & unrelaxed \\
\hline
(001) & h & 3.40 & 3.41 & 3.28 & 3.32 & 3.07 & 3.07 \\
      & b & 2.63 & 2.50 & 2.51 & 2.42 & 2.45 & 2.38 \\
      & t & 1.60 & 1.60 & 1.49 & 1.55 & 1.66 & 1.65 \\
\hline
(111) & h & 4.23 & 4.27 & 4.10 & 4.11 & 3.89 & 3.88 \\
      & b & 4.44 & 3.86 & 4.33 & 3.72 & 4.14 & 3.55 \\
      & t & 3.55 & 2.60 & 3.52 & 2.55 & 3.38 & 2.67 \\
\end{tabular}
\end{ruledtabular}
\end{table*}

The systems studied contain a benzene molecule connected to two Au leads through S, Se, or Te anchoring atoms.  While our main focus is on the chemical trends among these anchoring atoms, the structure of the contact is an important issue as well. Two different Au lead orientations, (001) and (111), are considered.  The in-plane size of the Au lead is set to be $2\sqrt{2} \!\times\! 2 \sqrt{2}$ for the (001) lead and 2$\times$2 for (111).  For checking the dependence on the in-plane lead size, 
we also use larger (001) leads, $3\sqrt{2} \!\times\! 3 \sqrt{2}$, and $4\sqrt{2} \!\times \! 4 \sqrt{2}$. 
Furthermore, we also use a (001)-($4\sqrt{2} \!\times \! 4 \sqrt{2}$)
periodic surface to model the (001) lead.
Because of the large separation between the molecule and its images (larger than 12{\AA}), 
this model will be a good approximation 
to the lead of an infinitely large surface. 
For the (111) lead, we consider different adsorption sites for the anchoring atom: hollow site (h), bridge site (b), and top site (t). From now on we use, for instance, the structural label S\_(001)\_h to denote the system with S anchoring atom adsorbed at the hollow site of the (001) lead surface.

The purpose here is to simulate possible experimental situations in break-junction experiments, in which different atomic structures may occur and, indeed, global structural equilibrium may not be reached.  In our calculation, the contact atomic relaxation is fully included by relaxing the molecule and the first two atomic layers of the lead surfaces, as well as the molecule-lead separation, while leaving the in-plane position of the anchoring atom fixed at the h, b, or t site.  The optimized atomic structure of the systems with an S anchoring atom are shown in Fig. 1; those for the other anchoring atoms are similar and therefore not shown. These fully optimized structures are called ``relaxed" structures hereafter (default if not specified), in contrast to ``unrelaxed" structures in which atoms in the leads are fixed at their bulk positions.  In the unrelaxed case, the molecule is fixed at its optimal isolated structure, the dangling bond on the S atom is saturated by an Au atom, and the distance between the S atom and the Au surface is optimized.

Because we use the bulk Au structure for the leads, the atomic relaxation consists of two parts: one is the relaxation of the bare Au lead with respect to its bulk structure, and the other is the relaxation of both the leads and the molecule induced by the molecule-lead interaction.  The latter is our primary interest; in fact, our calculations show that the former is very small, which is consistent with the very small surface relaxation of unreconstructed infinite Au surfaces.

For the electronic structure calculation, we use Siesta, an efficient full DFT package \cite{siesta}.  A high level double zeta plus polarization basis set (DZP) is adopted for all atomic species.  The PBE version of the generalized gradient approximation (GGA) \cite{pbe} is used for the electron exchange and correlation, and optimized Troullier-Martins pseudopotentials \cite{tmpp} are used for the atomic cores.  The atomic structure of the molecule, lead surfaces, and molecule-lead separation are fully optimized, with residual forces less than 0.02 eV/{\AA}.

For the transport calculation \cite{trank1} we divide an infinite LML system into three parts: left lead ($L$), right lead ($R$), and device region ($C$) which contains the molecule and large parts of the left and right leads, as shown in Fig. 1, so that the molecule-lead interactions can be fully accommodated. For a steady state situation in which the region $C$ is under a bias $V_b$ (zero or finite), its density matrix ($\mathbf{D}_C$) and Hamiltonian ($\mathbf{H}_C$) can be determined selfconsistently by the DFT+NEGF techniques \cite{datta1,datta2,pal1,mcdcal,transiesta,trank1}. The Kohn-Sham wave-functions are used to construct a single-particle Green function from which the transmission coefficient at any energy, $T(E,V_b)$, is calculated. The conductance, $G$, then follows from a Landauer-type relation.

%=========================== Fig. 2 =========================
\begin{figure}%[tbh]
\includegraphics[angle=  0,width= 7.5cm]{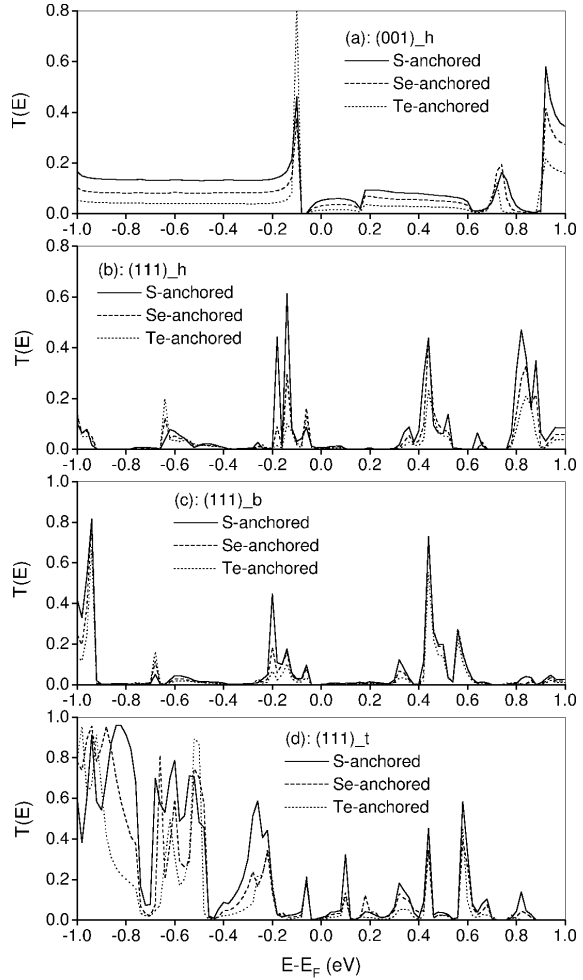}
\caption{Comparison among the transmission functions of the S-, Se-, and Te-anchored systems: (a) (001) with adsorption at the hollow site, (b) (111) hollow site, (c) (111) bridge, and (d) (111) top site. The conductance is largest in the (001) case, and decreases as the atomic number of the anchoring group increases.
} 
\end{figure}

\section{Results and Discussion}

\subsection{Structure}

We find that if the anchoring atom is adsorbed at the hollow site, the effect of contact atomic relaxation is minor and independent of the lead orientation [see Fig. 1 (a) and (b)].  However, if the anchoring atom is adsorbed at the b or t site [for a Au(111) lead], the relaxation effect becomes significant [Fig. 1 (c) and (d)].  This behavior is understandable because the h site is the bulk atomic position, so that the adsorption of an anchoring atom at it will not noticeably change the directional binding character of the surface. For the b or t site, in contrast, the directional binding character can be significantly modified by an adsorbed atom, leading to the significant contact atomic relaxation shown.

The relative stability of the different adsorption sites is related to the binding energy between the molecule and the lead, given by the difference in total energy between the whole system and the two subsystems using the same supercell and k-sampling.  Table I summarizes the results for both (001) and (111) leads connected by the three anchoring atoms, both relaxed and unrelaxed. We wish to point out three features: First, the binding energy on (111) is larger than that for (001). This is a straight forward result of the different directional binding character of the two surfaces with the group-VI anchoring atoms: the coordination on the (111) surface is smaller than on (001) and so is more favorable to the group-VI anchoring atom.  Second, for the Au(001) lead, the most stable adsorption site, regardless of contact relaxation, is the h site [shown in Fig. 1 (a)] followed by b and then t. The energy gains for b and t are quite large: for the relaxed S-anchored system, for instance, the energy increase for b and t sites are 0.77 and 1.80 eV/contact, respectively.  Third, for the Au(111) lead the situation is not so simple.  If the contact is not relaxed, the most stable adsorption site is h followed by b and t; however, after relaxation, the most stable site becomes b followed by h and t. The energy differences are smaller here compared to the (001) case: for the unrelaxed S-anchored system, the energy difference between h and b is 0.41 eV/contact while after relaxation it is only $-$0.21 eV/contact. 

Our results are largely in agreement with previous results concerning chemisorption on Au(111). Short alkane thiols on unrelaxed Au(111) prefer the h site \cite{franzen}. In a study of the chemisorption of 1,4-benzene-dithiol perpendicularly on the surface of an unrelaxed Au$_{25}$ cluster \cite{ricca}, the most stable adsorption site was h followed by b then t, as in our results. Finally, a systematic {\it ab initio} calculation of S-C$_6$H$_5$ chemisorbed on a Au$_{29}$ cluster \cite{weber}, in which the whole structure was fully relaxed in different geometrical configurations, it was found that several b sites have the lowest energy, consistent with our result for the relaxed S-anchored system. Despite this agreement with previous results, we caution that the structural optimization and energetics of the different structures in this paper may include a significant contribution from the small width of the Au leads; as a result, the present results may differ quantitatively from those for chemisorption on either an infinite surface or a small Au cluster. 

Considering that in break-junction experiments different contact atomic structures will occur and an adsorbed molecule may not be at its global equilibrium position, we investigate the three adsorption sites on Au(111)-leads.  For Au(001)-leads, however, we only investigate the most stable adsorption site (h) because it has a substantially lower energy than the next most stable one.

\subsection{Effects of structure on transmission}
% and different contact atomic configurations}

%============================== Table 2 ========================
\begin{table}[tb]
\caption{Equilibrium conductance ($G$, in units of $2e^2/h$) and molecule-lead electron transfer ($\Delta Q$, in units of electron charge). A positive $\Delta Q$ means that electrons are transferred from lead to molecule. `h-Au' inidcates that the molecule is connected to the leads via an additional Au atom adsorbed at the hollow site.  The sensitivity to both atomic structure and type of anchoring atom shows that treating the contact realistically is critical for reliable calculations in molecular electronics.
}
\begin{ruledtabular}
\begin{tabular}{ccccc}
anchoring  & lead         & adsorption &             &   \\
atom       & orientation  & site       & $\Delta$Q & G \\
\hline
S & (001) & h & $-$0.048 & 0.053 \\
% &   &  h-Au & +0.261 & 0.740 \\
  & (111) & h & +0.053 & 0.008 \\
  &       & b & +0.117 & 0.002 \\
  &       & t & +0.237 & 0.013 \\
  &   &  h-Au & +0.228 & 0.490 \\
  & jellium &   & +0.1$^a$\ \   & 0.036$^a$ \\
\hline
Se& (001) & h & $-$0.206 & 0.031 \\
% &   &  h-Au & +0.283 & 0.660 \\
  & (111) & h & $-$0.010 & 0.005 \\
  &       & b & +0.036 & 0.002 \\
  &       & t & +0.170 & 0.009 \\
  &   &  h-Au & +0.235 & 0.550 \\
  & jellium &   &        & 0.12$^a$ \\
\hline
Te& (001) & h & $-$0.294 & 0.014 \\
% &   &  h-Au & +0.260 & 0.420 \\
  & (111) & h & $-$0.044 & 0.002 \\
  &       & b & $-$0.002 & 0.001 \\
  &       & t & +0.167 & 0.004 \\
  &   &  h-Au & +0.240 & 0.430 \\
  & jellium &   & $-$0.5$^a$\ \   & 0.88$^a$ \\ 
\end{tabular}
\end{ruledtabular}
\centerline{$^a$Result from Ref.\cite{jellium}}
\end{table}

In Fig. 2 we show for the systems studied the transmission function $T(E)$ under zero bias.  The values of the equilibrium conductance are summarized in Table II, together with the molecule-lead electron transfer determined by a Mulliken population analysis.  
%something else ???

In Table II, first note that the Au(001) lead gives a much larger conductance (by about 6 times) than the (111) lead for all three anchoring atoms -- S, Se, and Te.  This can be understood by analyzing the molecule-lead coupling and electron transfer for the two lead orientations:  In the (001) case, the anchoring atom has four nearest neighour Au atoms while there are only three for (111). This difference in the contact atomic configuration certainly affects the molecule-lead coupling, which may also lead to a difference in the molecule-lead electron transfer.  However, for the S-anchored systems, both surfaces result in very small electron transfer (Table II), but their transmission functions are essentially different [Fig. 2 (a) and (c)].  In particular, there is a small peak around the Fermi energy for (001) which is absent for (111), indicating that the difference in equilibrium conductance is mainly due to the different molecule-lead couplings. An interesting aspect here is the difference between the binding (Table II) and the transport with the Au(001) and Au(111) leads: contacts with good (poor) binding have, however, relatively poor (good) transparency. This is an illustration that for completely different contact structures, binding and transport character are not necessarily related.

%=========================== Fig. 3 =========================
\begin{figure}[b]
\includegraphics[angle=  0,width= 7.0cm]{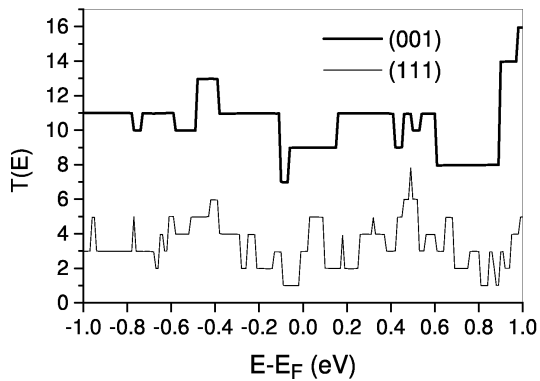}
\caption{Transmission functions of infinitely long wires: Au(001)-2$\sqrt{2} \!\times\! 2 \sqrt{2}$ (solid line) and Au(111)-$2 \!\times\! 2$ (thin line). Note the much larger average transmission for (001) and the large fluctuation for (111).
}
\end{figure}

Another difference between the two lead orientations is that the overall structure of $T(E)$ is totally different (Fig. 2): the $T(E)$ for Au(001) are quite smooth while those for Au(111) have many sharp features.  Possible reasons for the latter include a Au(111) lead that is too thin or its much lower symmetry.  Note that other calculations \cite{transiesta} also find sharp structure when using a Au(111) lead.  To dramatize the difference in transport between the two lead orientations, we calculate $T(E)$ for infinite pure (001) and (111) wires (Fig. 3). Because of the infinite periodic structures of the two wires, both $T(E)$  are step functions. As can be seen, the average transmission of the (001) wire is larger than that of the (111) wire, and the (111) transmission coefficient fluctuates strongly. These strong fluctuations are related to the sharp features in the transmission through the molecule.

Table II shows a clear trend in the equilibrium conductance for the different adsorption sites: $G (\mathrm{t}) >$ $G (\mathrm{h}) >$ $G(\mathrm{b})$ for all the S-, Se-, and Te-anchored (111) systems. Comparing to the binding energies for the different sites (Table I), we see again that there is no direct relation between transport and binding properties when comparing different structures. Note that the maximum difference in $G$ for the three sites can be up to about 6 times.  

In contrast, the trend in molecule-lead electron transfer is $\Delta Q(\mathrm{t}) >$ $\Delta Q(\mathrm{b}) >$ $\Delta Q(\mathrm{h})$.  Here we can also find cases where the electron transfer for different adsorption sites is very close but the conductance is quite different [in Table II compare Se\_(111)\_h to Se\_(111)\_b, or Te\_(111)\_h to Te\_(111)\_b].  These indicate that the trend in the equilibrium conductance is mainly due to differing molecule-lead coupling for the three adsorption sites rather than simply differing charge transfer.

This conclusion can also be reached from Fig. 2.  In Fig. 2 (b) and (c), because of the similar electron transfer (Table II), $T(E)$ for the h- and b-adsorbed systems are similar around the Fermi energy. In addition, the relative positions of the Fermi energy in the gap (which is comparable to the $\sim$ 0.3~eV HOMO-LUMO gap of the molecule including the two anchoring atoms) are similar.  However, there is a small shoulder at the Fermi energy for the h-adsorbed cases which is absent in the case of b-adsorption; this is the signature of the difference in the molecule-lead coupling.

%============================== Table 3 ========================
\begin{table}[t]
\caption{Current (in $\mu$A) for several systems
under 1~V and 3~V bias. The notation is as in Table II.
}
\begin{ruledtabular}
\begin{tabular}{ccccc}
anchoring  & lead         & adsorp. & \multicolumn{2}{c}{current} \\
atom       & orient.  & site       & 1 V & 3 V \\
\hline
S & (001) & h & 6.71 & 44.82 \\
  & (111) & b & 1.57 & 20.77 \\
  &   &  h-Au & 2.29 & \ 7.36 \\
\hline
Se& (001) & h & 4.74 & 32.57 \\
  & (111) & b & 1.29 & 14.81 \\
  &   &  h-Au & 2.37 & \ 6.40 \\
\hline
Te& (001) & h & 2.37 & 17.43 \\
  & (111) & b & 0.94 & \ 9.13 \\
  &   &  h-Au & 2.00 & \ 5.13 \\
\end{tabular}
\end{ruledtabular}
\end{table}

Results for finite bias provide further support for these conclusions. In Table III we list the current under 1 and 3~V biases for the (001)\_h and (111)\_b systems with each of the anchoring atoms.  The first thing we should notice is that the amplitude of the current for the S-anchored system is consistent with both a previous selfconsistent DFT+NEGF calculation using a cluster geometry \cite{xue1,xue2} and a self-consistent jellium model calculation \cite{jellium2}. All of these theoretical results are much larger than experimental results \cite{iexpt} 
%\marginpar{\bf More refs here?}
(as well as the result from a non-selfconsistent DFT+NEGF calculation \cite{derosa}). The reason for this discrepancy between theory and experiment is still an open problem; it may be related to the structural difference between the theoretical model and the real experimental conditions. 

%=========================== Fig. 4 =========================
\begin{figure*}[tbh]
\includegraphics[angle=  0,width= 6.8cm]{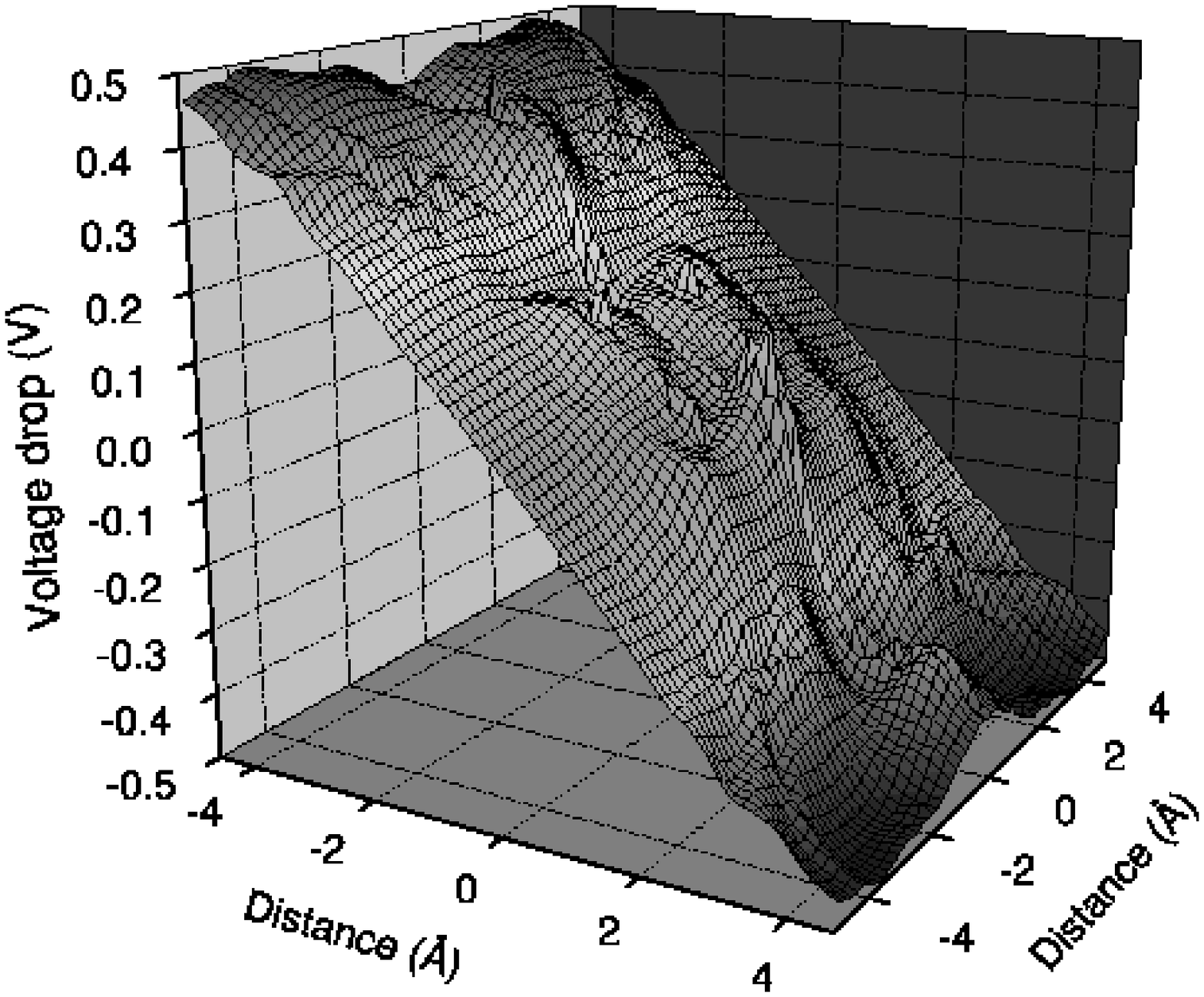} (a) 
\includegraphics[angle=  0,width= 6.8cm]{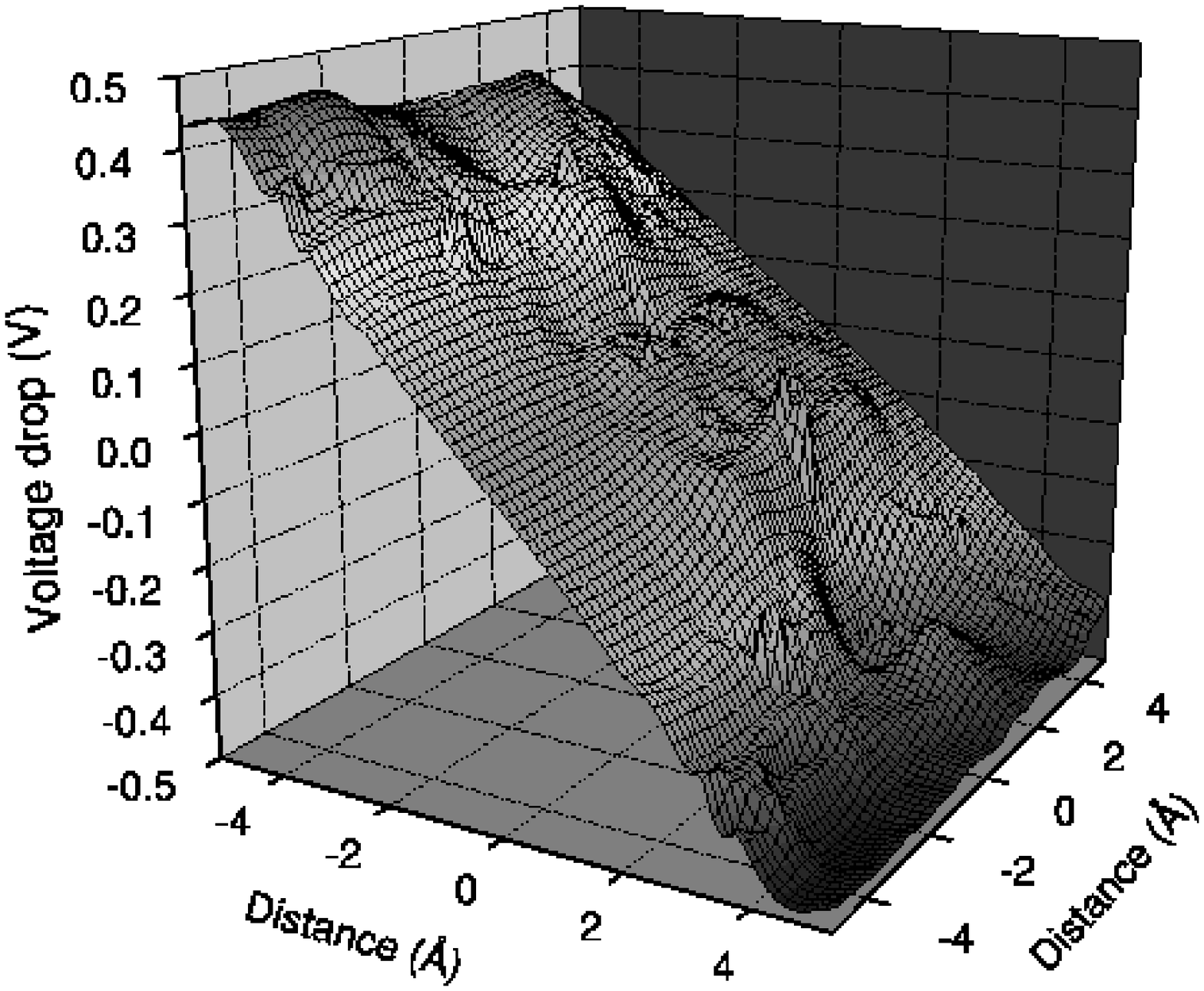} (d)
\includegraphics[angle=  0,width= 6.8cm]{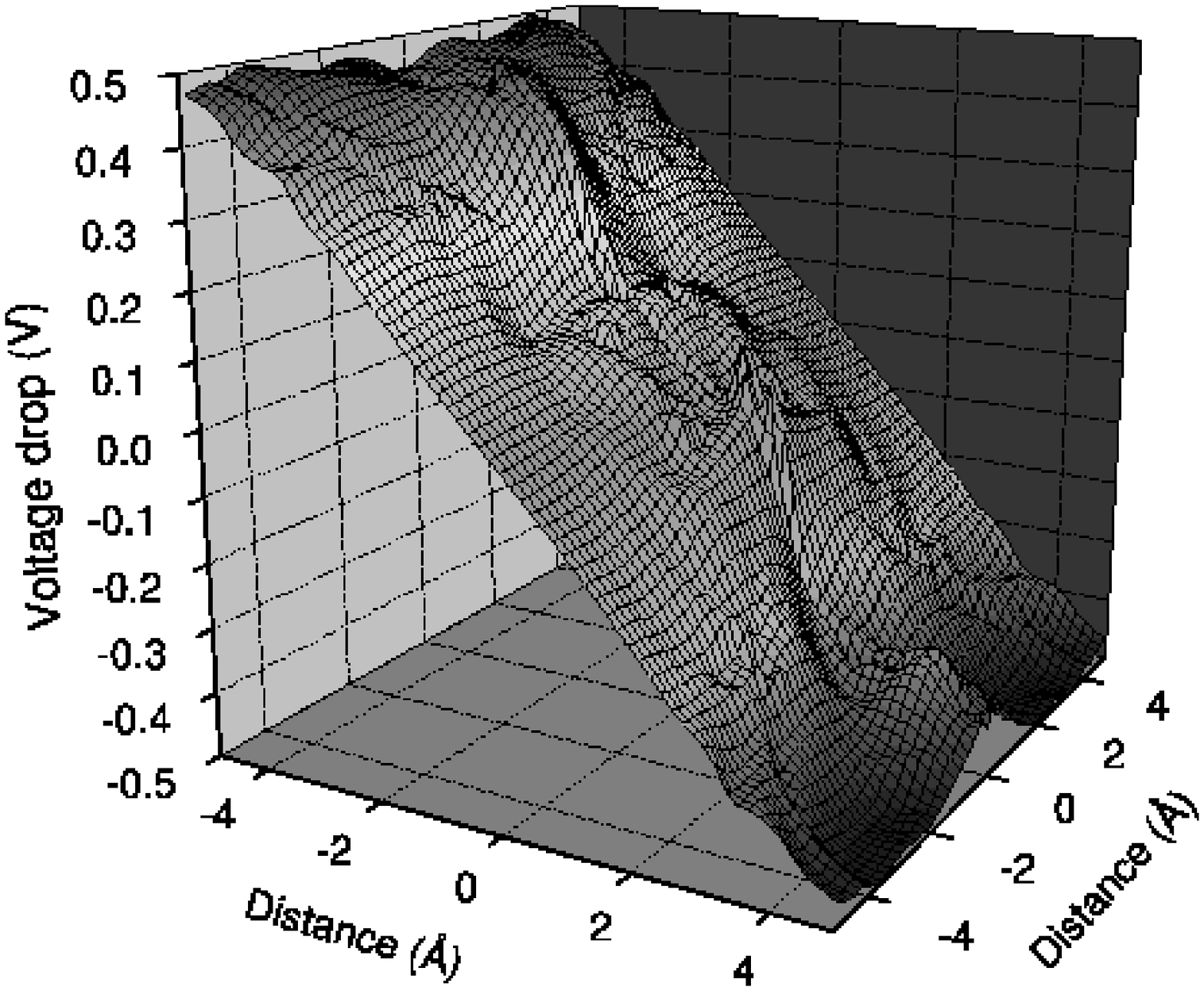} (b) 
\includegraphics[angle=  0,width= 6.8cm]{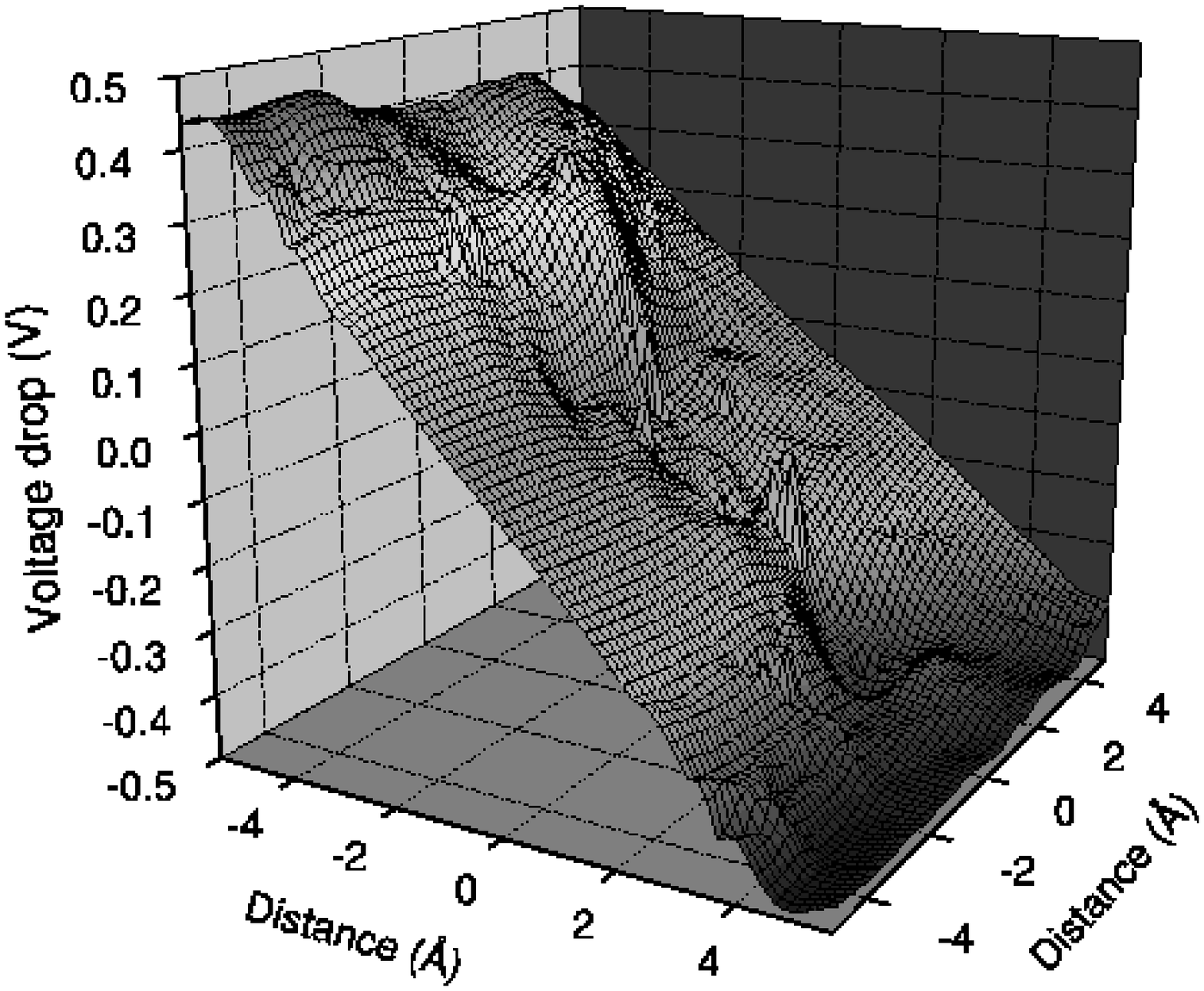} (e)
\includegraphics[angle=  0,width= 6.8cm]{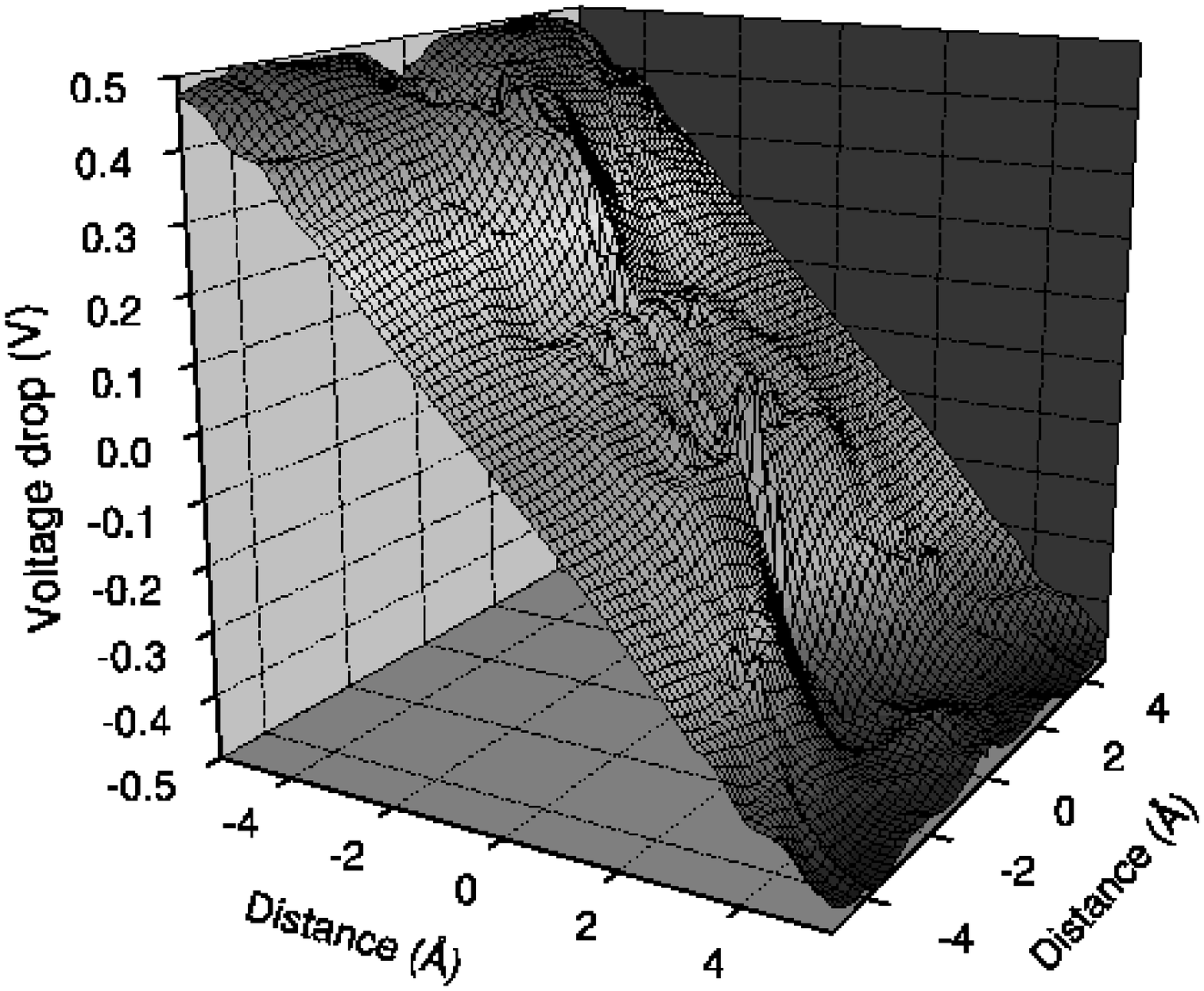} (c) 
\includegraphics[angle=  0,width= 6.8cm]{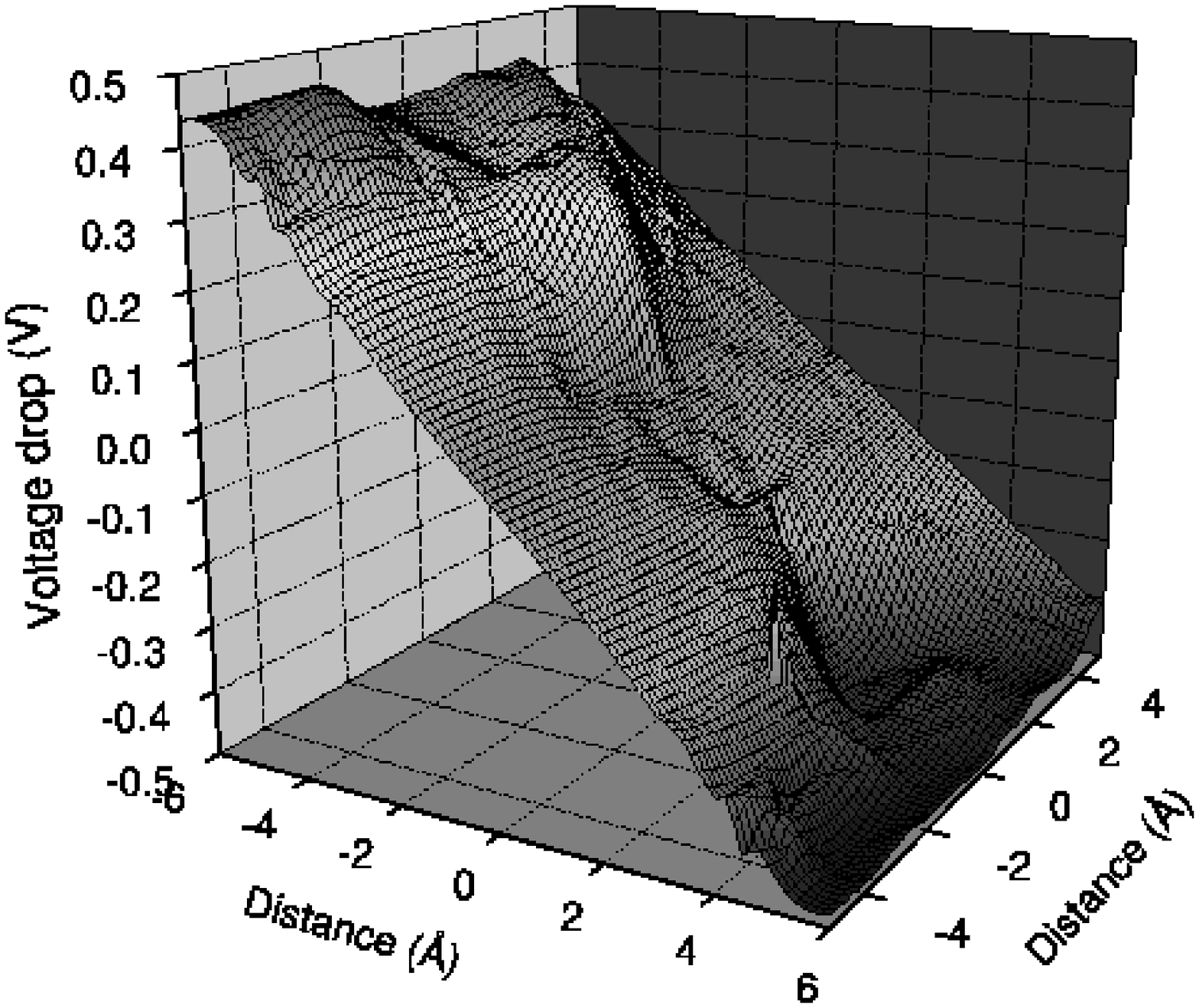} (f) 
\label{fig_vdmap001}
\caption{Voltage drops for 1 V bias (between the two lead surfaces as shown in Fig. 1). The contour surfaces are plotted in the plane crossing the benzene ring. The first column is for  (001)\_h cases: (a) S\_(001)\_h, (b) Se\_(001)\_h, and (c) Te\_(001)\_h. The second column is for (111)\_b cases: (d) S\_(111)\_b, (e) Se\_(111)\_b, and (f) Te\_(111)\_b. Note the similar voltage drop behavior for the different anchoring atoms and lead orientations.
}
\end{figure*}

Under finite bias the current for (001) is much larger than that for (111), although the relative difference tends to decrease with increasing bias. To show directly differences between the two lead orientations under a finite bias, the voltage drops for 1~V bias between the two lead surfaces are shown in Fig. 4 for the (001)\_h and (111)\_b cases.  Despite the difference in conductance between the two lead orientations, the voltage drops are strikingly similar. Note features in the middle corresponding to the benzene ring, features between the anchoring atom and the surface, and features between the anchoring atom and the ring. Looking carefully in Fig. 4 [(a) vs. (d), (b) vs. (e), and (d) vs. (f)], we see that the voltage drop between the left surface and the anchoring atom is slightly sharper in the (111) systems than in the (001) cases, indicating that the surface-anchoring-atom coupling is slightly weaker at the (111) contact than at the (001) contact.

\subsection{Chemical trends}

The chemical trends of different anchoring groups is a particularly interesting and important aspect of molecular electronics \cite{mol2,xue3}. They demonstrate, for instance, the critical role of realistic contact atomic structure in electron transport through molecules.  In Fig. 2 and more clearly in Table II, a chemical trend is evident in both the molecule-lead electron transfer and the equilibrium conductance for the three anchoring atoms we study, {\it regardless} of the lead orientation or adsorption site: As the atomic number of the anchoring atom increases, S $\rightarrow$ Se $\rightarrow$ Te, both the electron transfer from the leads to the molecule and the conductance decrease. 

Our result of decreasing $G$ with increasing atomic number is directly opposite to the conclusion reached by a previous calculation \cite{jellium} using the jellium model for the Au leads (Table II).  Besides the qualitative difference in the chemical trend of conductance, the present calculation shows that the maximum difference in equilibrium conductance is less than 5 times in contrast with 25 times found in the previously.  The chemical trend in the electron transfer is consistent with the electronegativities of the three species [2.58, 2.55, and 2.1, respectively (Pauling scale)]; it is also consistent with the trend in binding energy (Table I), indicating that for similar contact structures the transport and binding properties \textit{are} related.  

%============================== Table 4 ========================
\begin{table}[t]
\caption{Equilibrium conductance (in units of $2e^2/h$) for the unrelaxed systems with larger Au(001) leads: $3\sqrt{2}\times 3 \sqrt{2}$, $4\sqrt{2}\times 4 \sqrt{2}$, and $4\sqrt{2}\times 4 \sqrt{2}$ periodic surface. Note the same chemical trend for the three anchoring atomsas in the case of the smaller 2$\sqrt{2}\times$2$\sqrt{2}$ leads (Table II).
}
\begin{ruledtabular}
\begin{tabular}{cccc}
anchoring & \multicolumn{3}{c}{lead size} \\
atom      & $3\sqrt{2}\times 3 \sqrt{2}$ & $4\sqrt{2}\times 4 \sqrt{2}$ & $4\sqrt{2}\times 4 \sqrt{2}$ surface \\
\hline
S  & 0.110 & 0.089 & 0.108 \\
Se & 0.076 & 0.064 & 0.087 \\
Te & 0.044 & 0.037 & 0.053 \\
\end{tabular}
\end{ruledtabular}
\end{table}

In order to check whether the clear chemical trend in $G$ is affected by the width of the leads, we carry out calculations for (001)\_h systems using two wider leads: $3\sqrt{2}\times 3\sqrt{2}$ and $4\sqrt{2}\times 4\sqrt{2}$. Furthermore, we also carry out test calculations using a (001)-($4\sqrt{2}\times 4\sqrt{2}$)
periodic surface for the leads, which will be a good approximation to an infinitely large surface
because of the large separation between the molecule and its images (larger than 12{\AA}).
To avoid prohibitive computational effort, we adopted a single zeta plus polarization (SZP) basis set here. Table IV shows that the chemical trend remains with the wider Au leads. 

%=========================== Fig. 5 =========================
\begin{figure}[tbh]
\includegraphics[angle=-90,width= 7.0cm]{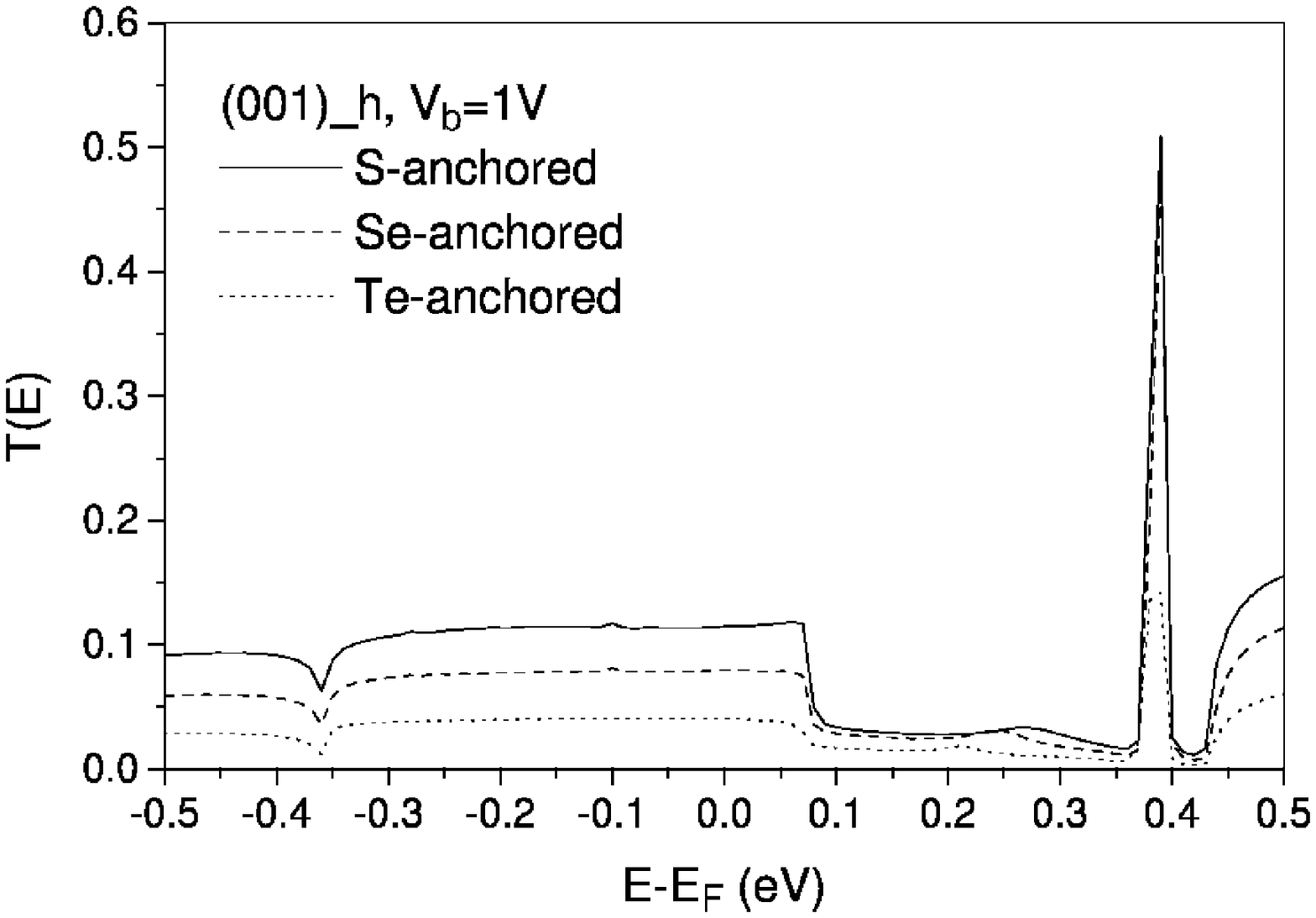} (a) \\
\includegraphics[angle=  0,width= 7.0cm]{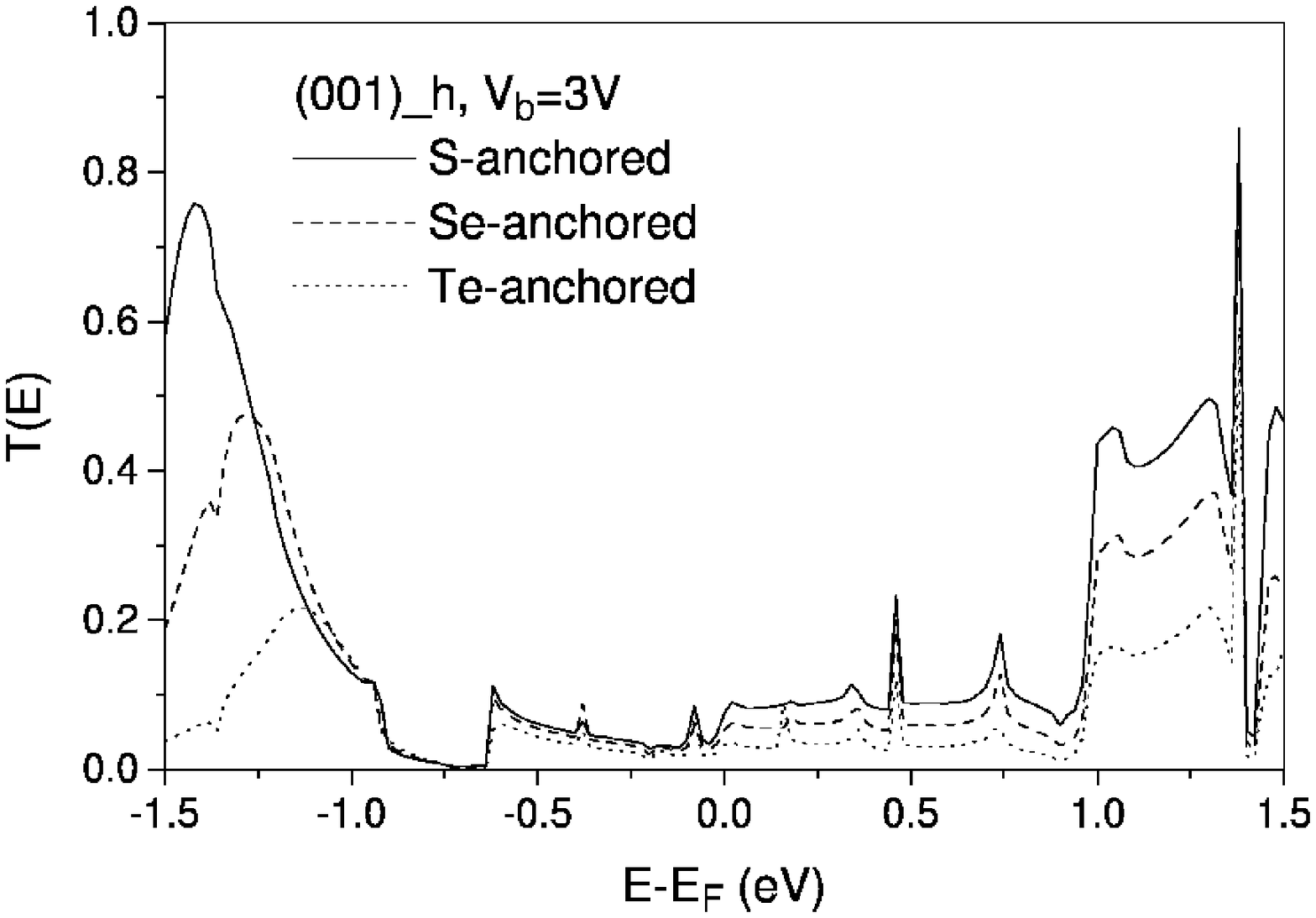} (b)
\caption{Comparison among the transmission functions for S-, Se-, and Te-anchoring in the case of (001) adsorption at the hollow site: (a) 1~V bias, (b) 3~V bias. Note that the transmission coefficient decreases as the atomic number of the anchoring group increases, the same trend as for zero bias (see Fig. 2).
}
\end{figure}
                                                                                                              
Results for finite bias provide further support for our conclusions.  Under 1 or 3~V bias, the current decreases with increasing atomic number of the anchoring atom for both (001) and (111) leads (Table III). $T(E)$, shown in Fig. 5 for the (001)\_h cases, decreases at most energies as atomic number increases. From the voltage drop profiles in Fig. 4, for both (001) and (111) the drops around the anchoring atoms become slightly sharper as the atomic number of the anchoring atom increases. This is direct evidence that the surface to anchoring-atom coupling under finite bias becomes weaker as atomic number increases.

In break junction experiments, the break surfaces are certainly not as flat as assumed above but rather very likely have atomic fluctuation (atomic scale roughness).  To see whether or not the clear chemical trend of the different anchoring atoms is affected by the change in contact atomic configuration, we consider an additional Au atom adsorbed at the h site of the two Au(111) lead surfaces. There has been previous work on this system using DFT+NEGF calculations \cite{xue1,xue2,bauschlicher} using an unrelaxed S\_(111)\_h configuration, as well as a calculation modeling the leads with jellium \cite{jellium2}. First, using an approximate non-self-consistent approach in which a constant imaginary self-energy is adopted for the Au lead, one study found that the current under 4~V bias depends strongly on the adsorption site on the Au lead \cite{bauschlicher}; when the connection was made by an apex Au atom, the current was significantly reduced. Second, in a systematic calculation for the unrelaxed S\_(111)\_h system adopting a cluster method in which only 6 Au atoms are included to form the device region \cite{xue1,xue2}, the equilibrium conductance was significantly increased by the addition of the Au atoms at each contact, but the current under large bias was decreased. The result for equilibrium conductance was opposite to that obtained with jellium leads \cite{jellium2}, showing that one must use caution when considering the latter.

In view of these previous results, our goal here is two fold: On the one hand, we wish to study this system with the more recently developed presumably more accurate methods; in particular, we will include 36 Au atoms per (111) lead and 50 per (001) lead in contrast to the 6 used previously. Summarizing the results below, we find that the more accurate methods give qualitatively the same results as those used previously. 

%=========================== Fig. 6 =========================
\begin{figure}[t]
\includegraphics[angle=  0,width= 6.8cm]{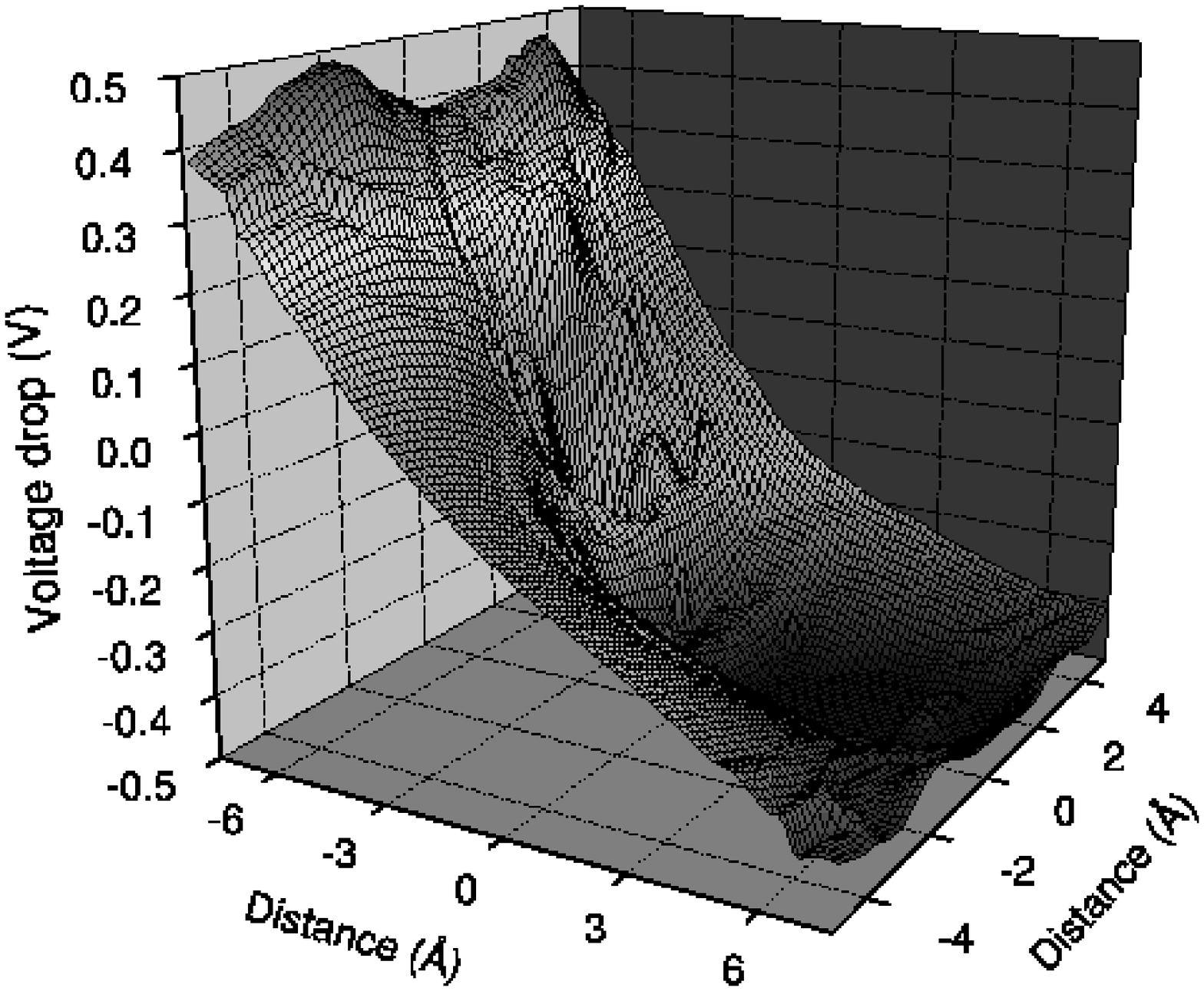} (a) \\
\includegraphics[angle=  0,width= 6.8cm]{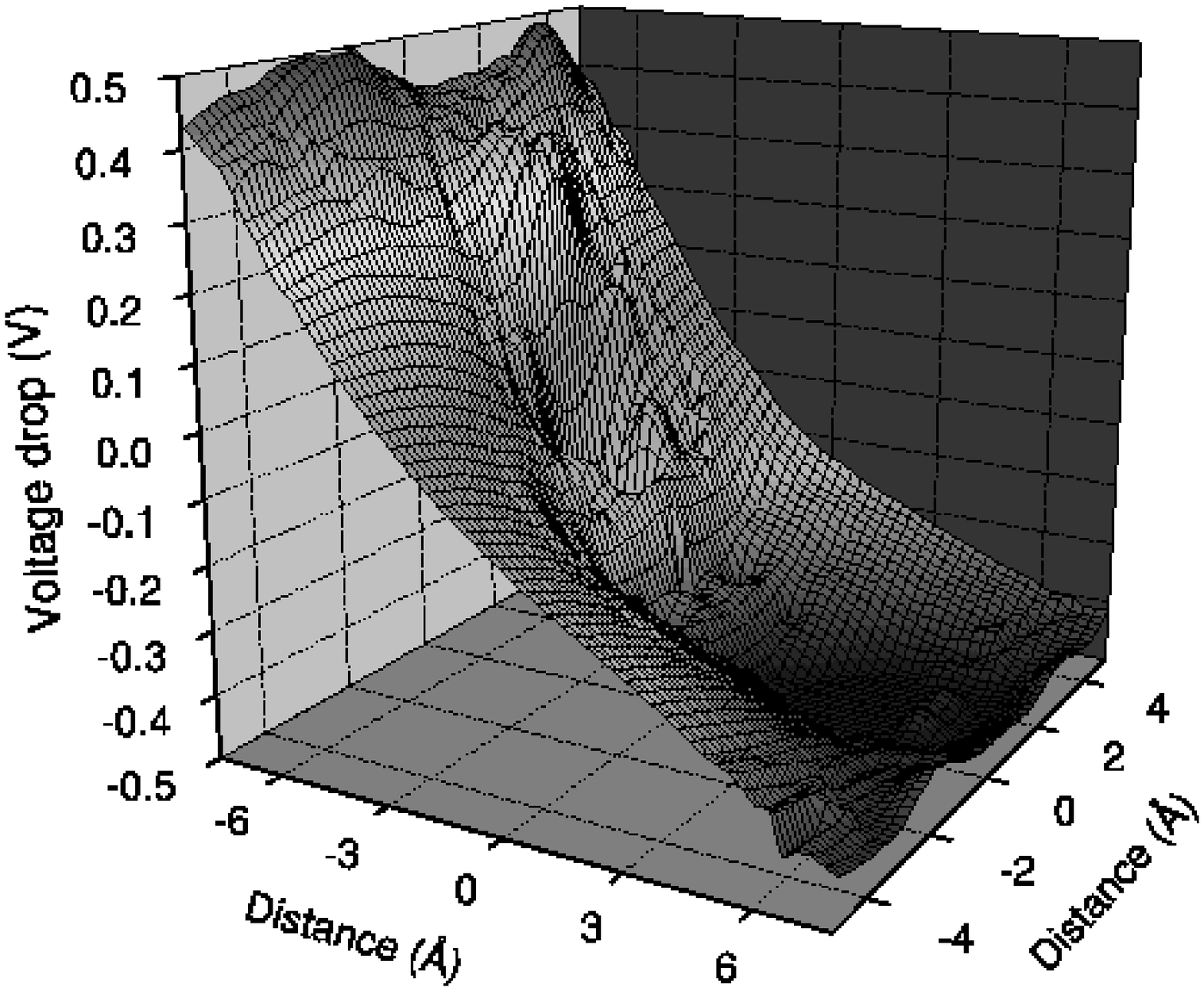} (b) \\
\includegraphics[angle=  0,width= 6.8cm]{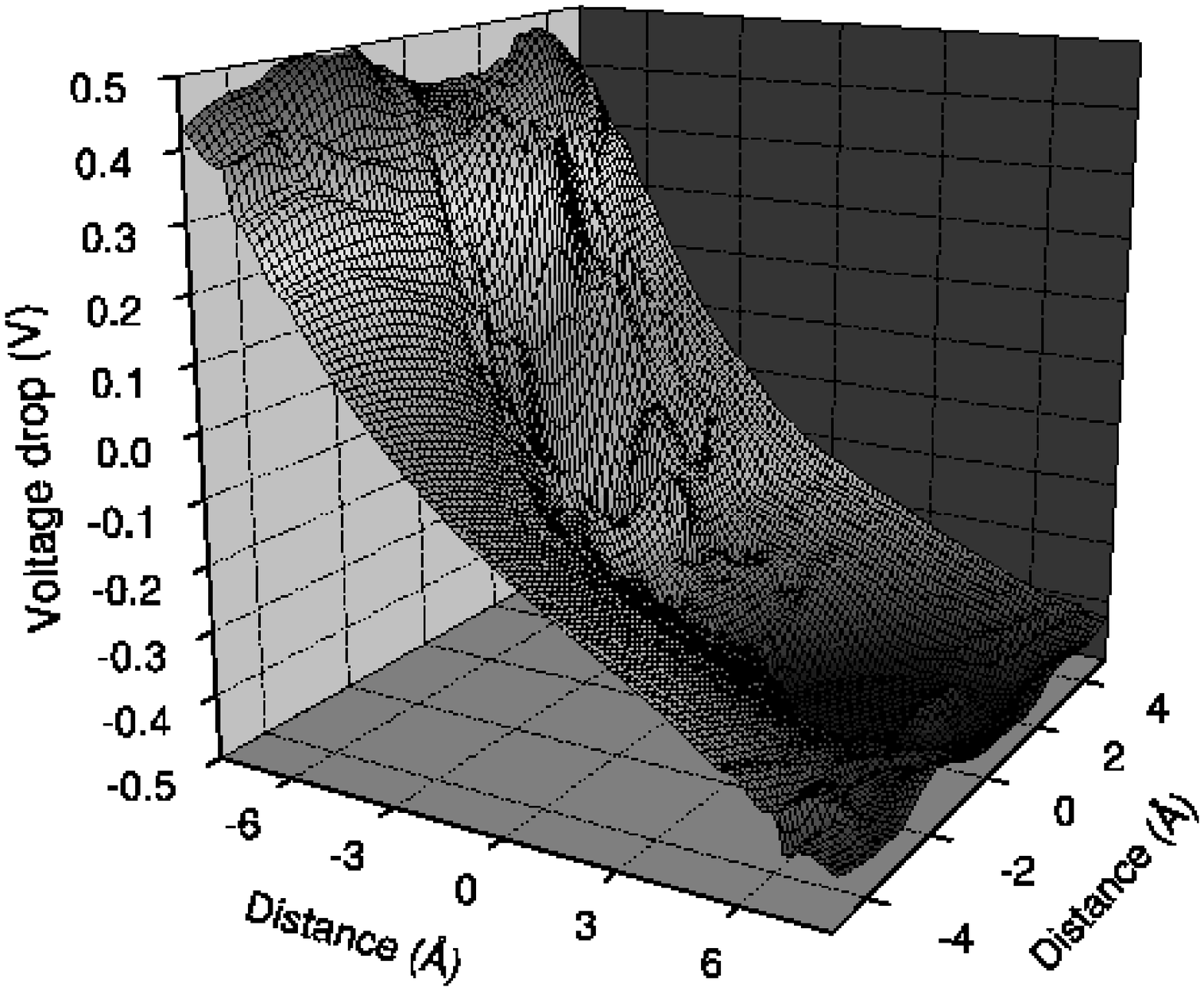} (c)
\label{fig_vdmap111}
\caption{Voltage drops for 1 V bias with an additional Au atom at each contact in the h site on (111): (a) S (b) S,and (c) Te. The additional Au atom is included as a part of the molecule not as a part of the surface. The contour surfaces are plotted in the plane crossing the benzene ring. Note that the voltage drop behavior is very similar for the three different anchoring atoms but very different from that in the systems without the additional Au atoms (see Fig. 4). 
}
\end{figure}

On the other hand, and more importantly, we wish to investigate the chemical trend in the effect of the additional Au atom. Our results for electron transfer and equilibrium conductance are listed in Table II, with the adsorption site denoted `h-Au'.  The additional Au atom increases significantly the electron transfer from the leads to the molecule, while the difference among the three anchoring atoms is strongly reduced. Because of the large electron transfer, the LUMO resonance becomes close to the Fermi energy; consequently, the equilibrium conductance also increases significantly. Note that introduction of the additional Au atoms at the contacts changes the chemical trend: Se becomes the best anchoring atom in terms of contact transparency.

Under a finite bias, the relatively large amount of charge transferred to the molecule causes a large density of states at energies between the left and right Fermi energies. This in turn causes very different voltage drop behavior in the presence of additional Au atoms: fluctuations in the voltage drop profile become large because of a large bias-induced polarization (compare Fig. 6 with Fig. 4). If the bias is relatively small (1~V in Table III), Se remains the best anchoring atom for contact transparency; however, if the bias is large (3~V), then the chemical trend reverts to S being best.  Although the introduction of the additional Au atoms increases significantly the molecular conductance for zero or small bias, it decreases the current under large bias (Table III).

\section{Summary}

By using a state-of-the-art {\it ab initio} method for electronic structure and electron transport, we have carried out a systematic calculation for the molecular conductance of benzene sandwiched between two Au electrodes. Our calculation is the first to fully include the effects of relaxing the contact atomic structure. The main results are: (1)~Detailed contact structure strongly affects molecule-lead coupling, electron transfer, and molecular conductance. (2)~There is no general relation between the binding strength of an anchoring atom and the transparency of the contact; however, in the special case of comparing contacts with very similar atomic structure, stronger binding does imply increased transparency. (3)~For ideally flat break surfaces, the equilibrium conductance decreases with increasing atomic number of the anchoring group regardless of the adsorption site, lead orientation, or bias. (4)~This chemical trend is, however, affected by the local contact atomic configuration: An additional Au atom at the contact with the Au(111) lead changes the best anchoring atom for small bias (from S to Se) although not for large bias. These results demonstrate the critical role of realistic contact atomic structure in electron transport through molecules.

\acknowledgments
This work was supported in part by the NSF (DMR-0103003).

\end{document}